\documentclass[12pt]{article}
\usepackage{subfloat}
\pdfoutput = 1
\begin{document}
\date{Today}
\title{{\bf{\Large  Analytic study of properties of holographic $p$-wave superconductors }}}

\author{
{\bf {\normalsize Sunandan Gangopadhyay}$^{a,c}
$\thanks{sunandan.gangopadhyay@gmail.com, sunandan@iucaa.ernet.in, sunandan@bose.res.in}},
{\bf {\normalsize Dibakar Roychowdhury}
$^{b,}$\thanks{dibakar@bose.res.in, dibakarphys@gmail.com}}\\
$^{a}$ {\normalsize Department of Physics, West Bengal State University, Barasat, India}\\
$^{b}$ {\normalsize  S.N. Bose National Centre for Basic Sciences,}\\{\normalsize JD Block, 
Sector III, Salt Lake, Kolkata 700098, India}\\[0.2cm]
$^{c}${\normalsize Visiting Associate in Inter University Centre for Astronomy \& Astrophysics,}\\
{\normalsize Pune, India}\\[0.1cm]
}
\date{}

\maketitle

\begin{abstract}
{\noindent In this paper, we analytically investigate the properties of $p$-wave holographic superconductors in $AdS_{4}$-Schwarzschild
background by two approaches, one based on the Sturm-Liouville eigenvalue problem and the other based on the matching of the solutions to the field equations near the horizon and near the asymptotic $AdS$ region. The relation between 
the critical temperature and the charge density has been obtained and the
dependence of the expectation value of the condensation operator on the temperature has been found.
Our results are in very good agreement with the existing numerical results. 
The critical exponent of the condensation also comes out to be $ 1/2 $  which is the universal value in the mean field theory.
  }
\end{abstract}
\vskip 1cm

\section{Introduction}
The first successful microscopic theory of superconductivity, BCS theory \cite{bcs}, was formulated
over fifty years ago and correctly describes the superconducting phenomenology 
of a large number of metals and alloys \cite{parks}. 
The basic idea of superconductivity in these weakly coupled systems is the spontaneous breaking at low temperatures of a $U(1)$ symmetry
due to a charged condensate. The condensate is a Cooper pair of electrons bound together by lattice vibrations or phonons.
However, it has been appreciated that the understanding of the pairing mechanism leading to the charged condensate 
is far from reality for materials exhibiting superconductivity at high temperatures (high $ T_c $ cuprates).
There are indications that the relevant new physics is strongly coupled and hence requires a new theoretical insight. 
One such insight comes from the $AdS$/CFT correspondence.

The AdS/CFT correspondence, which has been a powerful tool to deal with strongly coupled systems, is regarded as the most remarkable
discovery in string theory. It provides an exact correspondence
between a gravity theory in a $(d+1) $ dimensional $AdS$ spacetime 
and a conformal field theory (CFT) living on its $d$-dimensional boundary \cite{adscft1}-\cite{adscft4}. In recent years the $AdS$/CFT correspondence has provided some remarkable theoretical insights in order to understand the physics of high $T_c$ superconductors. The holographic description of  $s$-wave superconductors consists of a AdS black hole and a complex scalar field minimally coupled to an abelian gauge field. The black hole admits scalar hair formation below certain critical temperature ($T_c$) indicating the onset of a charged scalar condensate in the dual CFTs. The mechanism behind this condensation is the breaking of a local $U(1)$ symmetry near the event horizon of the black hole \cite{hs1}-\cite{hs6}.

Examples of superconducting black holes exhibiting $p$-wave gap has also been found \cite{hs13}-\cite{hs13a}. The results here are based
on classical solutions to field equations of Einstein-Yang-Mills theory with a negative cosmological constant
\begin{eqnarray}
S=\frac{1}{2\zeta^2}\int d^{4}x~\left\{R-\frac{1}{4}(F^{a}_{\mu\nu})^2 +\frac{6}{l^2}\right\}
\label{intro}
\end{eqnarray}
where $F^{a}_{\mu\nu}$ is the field strength of an $SU(2)$ gauge field. Here the idea is to consider
a $U(1)$ subgroup as the gauge group of electromagnetism and to persuade the 
gauge bosons, charged under this $U(1)$, to condense outside the horizon of the black hole.

Till date a number of numerical as well as analytical studies have been performed on both $s$ and $p$-wave holographic superconductor models 
\cite{hs6}-\cite{hs24}. However, analytical study of properies of holographic $p$-wave superconductors in $AdS_4$-Schwarzschild
background has so far been missing in the literature. 
In this paper, we present two analytical approaches, one based on the Sturm-Liouville (SL) eigenvalue problem \cite{hs7}
and the other based on the matching of the solutions to the field equations near the horizon and near the asymptotic $AdS$ region \cite{hs8} 
to find the relation between the critical temperature and the charge density and to compute the expectation value 
of the condensation operator for holographic $p$-wave superconductors.  This eventually helps us to compare the strength of  both the methods simultaneously.
The critical exponent for the condensation near the critical temperature can also be obtained 
naturally in our analysis. We compare our analytical results with the numerical results  existing 
in the literature\cite{hs13}-\cite{hs13a}. 
It is reassuring to note that all our calculations have been carried out in the probe limit.      

This paper is organized as follows. In section 2, we provide the basic holographic set up for the $p$-wave superconductors,
considering the background of a Schwarzschild-$AdS_4$ spacetime.
In section 3, ignoring the back reaction of the dynamical matter field on the space time metric, we compute the critical temperature as well as the temperature dependence of the condensate using the SL eigenvalue problem. In section 4, we present a simple analytical
method based on matching the solutions to the field equations near the horizon and near the asymptotic $AdS$ region
to find the relation between the critical temperature and the charge density and to compute the expectation value 
of the condensation operator and compare our results with the SL method and the numerical results. We conclude finally in section 5.

\section{Basic set up for $p$-wave superconductors}
Our construction of the holographic $p$-wave superconductor is based on the fixed
background of Schwarzschild-AdS space time.
The metric of a planar Schwarzschild-AdS black hole reads
\begin{eqnarray}
ds^2=-f(r)dt^2+\frac{1}{f(r)}dr^2+r^2(dx^2+dy^2)
\label{m1}
\end{eqnarray}
with
\begin{eqnarray}
f(r)=r^{2}-\frac{r_{+}^3}{r}
\label{metric}
\end{eqnarray}
in units in which the AdS radius is unity, i.e. $l=1$.
The Hawking temperature is related to the horizon radius ($r_+$) and is given by
\begin{eqnarray}
T=\frac{3r_+}{4\pi}~.
\label{temp}
\end{eqnarray}
In order to study the holographic $p$-wave superconductors in the probe limit, we need to
introduce an $SU(2)$ Yang-Mills action in the bulk theory. The Lagrangian density of this is given by
\begin{eqnarray}
\mathcal{L}=-\frac{1}{4}F^{a}_{\mu\nu}F^{a\mu\nu}
\label{lagrangian}
\end{eqnarray}
where 
\begin{eqnarray}
F^{a}_{\mu\nu}=\partial_{\mu}A^{a}_{\nu}-\partial_{\nu}A^{a}_{\mu}+\epsilon^{abc}A^{b}_{\mu}A^{c}_{\nu}
\label{fst}
\end{eqnarray}
is the Yang-Mills field strength, $(a,b,c)=(1,2,3)$ are the indices of the generators of $SU(2)$ algebra. 
$A^{a}_{\mu}$ are the components of the mixed valued gauge field $A=A^{a}_{\mu}\tau^{a}dx^{\mu}$, where 
$\tau^{a}$ are the $SU(2)$ generators satisfying the usual commutation relations.

\noindent We now choose the following ansatz for the gauge field \cite{hs13, hs13b}
\begin{eqnarray}
A=\phi(r)\tau^{3}dt +\psi(r)\tau^{1}dx~.
\label{vector}
\end{eqnarray}
In the above ansatz, the $U(1)$ subgroup of $SU(2)$ generated by $\tau^3$ is identified to be
the electromagnetic $U(1)$ \cite{hs13, hs13b}. The gauge boson with non-zero component $\psi(r)$ along
$x$-direction is charged under $A^{3}_{t}=\phi(r)$. 
According to the AdS/CFT dictionary, $\phi(r)$ is dual to
the chemical potential in the boundary field theory while $\psi(r)$ is dual to the $x$-component
of some charged vector operator $J$. The condensation of $\psi(r)$ spontaneously breaks the
$U(1)$ gauge symmetry and triggers the phenomena of superconductivity on the boundary
field theory.

\noindent The equations of motion for the fields $\phi(r)$ and $\psi(r)$ computed on this ansatz read
\begin{eqnarray}
\label{e1}
\phi''(r)+\frac{2}{r}\phi'(r)
-\frac{\psi^2 (r)}{r^{2}f}\phi(r)=0\\
\psi^{''}(r)+\frac{f'}{f}\psi'(r)
+\frac{\phi^{2}(r)}{f^2}\psi(r)=0
\label{e2}
\end{eqnarray}
where prime denotes derivative with respect to $r$. In order to solve the
non-linear equations (\ref{e1}) and (\ref{e2}), we need
to seek the boundary condition for $\phi(r)$ and $\psi(r)$ near the black
hole horizon $r\sim r_+$ and at the spatial infinity
$r\rightarrow\infty$. At the horizon, we require $\phi(r_+)=0$ for the $U(1)$ gauge field
to have a finite norm and $\psi(r_{+})$ should be finite. Near the boundary of the bulk, we have
\begin{eqnarray}
\label{bound1}
\phi(r)=\mu-\frac{\rho}{r}\\
\psi(r)=\frac{\psi^{(1)}}{r}~.
\label{bound2}
\end{eqnarray}
$\mu$ and $\rho$ are dual to the chemical potential and charge density of the boundary CFT
and $\psi^{(1)}$ is dual to the expectation value of the condensation operator $J$ at the boundary.

\noindent Under the change of coordinates $z=\frac{r_{+}}{r}$,  the field equations become
\begin{eqnarray}
\label{e1aa}
\phi''(z)-\frac{\psi^2 (z)}{r_{+}^{2} (1-z^3)}\phi(z) =0\\
\psi''(z)-\frac{3z^2}{1-z^3}\psi'(z)
+\frac{\phi^{2}(z)}{r_{+}^{2}(1-z^3)^2}\psi(z)=0
\label{e1a}
\end{eqnarray}
where prime now denotes derivative with respect to $z$. These equations are to be solved in the
interval $(0, 1)$, where $z=1$ is the horizon and $z=0$ is the boundary.
The boundary condition $\phi(r_+)=0$ now becomes $\phi(z=1)=0$.

\section{Sturm-Liouville method}
\subsection{Relation between critical temperature and charge density}
With the above set up in place, we now move on to investigate the relation between the critical temperature and the charge density. 

\noindent At the critical temperature $T_c$, $\psi=0$, so the field equation (\ref{e1aa})
reduces to
\begin{eqnarray}
\phi''(z)=0.
\label{e1b}
\end{eqnarray}
With the boundary condition (\ref{bound1}), the solution of this equation reads
\begin{eqnarray}
\phi(z)=\lambda r_{+(c)}(1-z)
\label{sol}
\end{eqnarray}
where 
\begin{eqnarray}
\lambda=\frac{\rho}{r_{+(c)}^2}~.
\label{lam}
\end{eqnarray}
 

\noindent Using the above solution, we find that as $T\rightarrow T_c$, 
the equation for the field $\psi$ approaches the limit 
\begin{eqnarray}
-\psi''(z)+\frac{3z^2}{1-z^3}\psi'(z)=\frac{\lambda^2 }{(1+z+z^2)^2}\psi(z)~.
\label{e001}
\end{eqnarray}

\noindent Near the boundary, we define \cite{hs7}
\begin{eqnarray}
\psi(z)=\frac{\langle J\rangle}{\sqrt 2 r_+} zF(z)
\label{sol1}
\end{eqnarray}
where $F(0)=1$.
Substituting this form of $\psi(z)$ in eq.(\ref{e001}), we obtain
\begin{eqnarray}
- F''(z) + \left(\frac{3z^2}{1-z^3}-\frac{2}{z}\right)F'(z) + \frac{3z}{1-z^3}F(z)
&=&\frac{\lambda^2 }{(1+z+z^2)^2}F(z) 
\label{eq5b}
\end{eqnarray}
to be solved subject to the boundary condition $F' (0)=0$. 

\noindent The above equation can be put in the Sturm-Liouville form 
\begin{eqnarray}
\frac{d}{dz}\left\{p(z)F'(z)\right\}-q(z)F(z)+\lambda r(z)F(z)=0
\label{sturm}
\end{eqnarray}
with 
\begin{eqnarray}
p(z)=z^{2}(1-z^3)~,~ q(z)=3z^3~,~r(z)=\frac{z^{2}(1-z)}{1+z+z^2}. 
\label{i1}
\end{eqnarray}
With the above identification, we now write down the eigenvalue $\lambda^2$ which minimizes the expression 
\begin{eqnarray}
\lambda^2 &=& \frac{\int_0^1 dz\ \{p(z)[F'(z)]^2 + q(z)[F(z)]^2 \} }
{\int_0^1 dz \ r(z)[F(z)]^2}\nonumber\\
&=&\frac{\int_0^1 dz\ \{z^{2}(1-z^3)[F'(z)]^2 + 3z^{3}[F(z)]^2 \} }
{\int_0^1 dz \ \frac{z^{2}(1-z)}{1+z+z^2}[F(z)]^2}~.
\label{eq5abc}
\end{eqnarray}
To estimate it, we use the following trial function
\begin{eqnarray}
F= F_\alpha (z) \equiv 1 - \alpha z^2
\label{eq50}
\end{eqnarray}
which satisfies the conditions $F(0)=1$ and $F'(0)=0$.

\noindent Hence, we obtain
\begin{eqnarray}
\lambda_\alpha^2 = \frac{60(\alpha-\frac{3}{4}-\frac{27\alpha^2}{40})}{(-130+21\alpha-10\sqrt{3}\pi\alpha
+ 30 (\alpha+4)\ln3)\alpha + 10(-9+\sqrt{3}\pi +3\ln3)} 
\end{eqnarray}
which attains its minimum at $\alpha \approx 0.5078$. 
The critical temperature therefore reads 
\begin{eqnarray}
T_c = \frac{3}{4\pi} r_{+(c)} = \frac{3}{4\pi} \sqrt{\frac{\rho}{\lambda_{\alpha=0.5078}}}\approx 0.1239\sqrt\rho 
\label{eqTc}
\end{eqnarray}
which is in very good agreement with the exact $T_c = 0.125\sqrt\rho$ \cite{hs13}.


\subsection{Critical exponent and condensation values}
In this section, we shall compute the condensation values of the condensation operator $J$ in the boundary field theory.

\noindent Away from (but close to) the critical temperature $T_c$, 
the field equation (\ref{e1aa}) for $\phi$ becomes (using eq.(\ref{sol1}))
\begin{eqnarray}
\label{aw1}
\phi''(z) &=&\frac{\langle J\rangle^2}{r_+^{2}}\mathcal{B}(z)\phi(z)\\
\mathcal{B}(z)&=&\frac{z^2 F^{2}(z)}{2r_+^{2}(1-z^3)}\nonumber
\end{eqnarray}
where the parameter $\langle J\rangle^2/r_+^{2}$ is small.
Now we expand $\phi(z)$ in the small parameter $\langle J\rangle^2/r_+^{2}$ as
\begin{eqnarray}
\frac{\Phi}{r_+}=\lambda (1-z)+ \frac{\langle J\rangle^2}{r_+^{3}} \chi(z) +\dots
\label{aw2} 
\end{eqnarray}
From eq(s)(\ref{aw1}, \ref{aw2}), we obtain the equation for
the correction $\chi(z)$ near the critical temperature
\begin{eqnarray}
\chi''(z) = \lambda r_{+}(1-z)\mathcal{B}(z)
\label{aw3}
\end{eqnarray}
with $\chi(1)=0=\chi'(1)$. 
Integrating both sides of the above equation between $z=0$ to $z=1$, we obtain
\begin{eqnarray}
\chi'(0)=-\lambda r_{+}\int_{0}^{1}(1-z)\mathcal{B}(z)dz~.
\label{aw5}
\end{eqnarray}
Now from eq(s)(\ref{bound1}, \ref{aw2}), we have
\begin{eqnarray}
\frac{\mu}{r_+}-\frac{\rho}{r_{+}^2}z&=&\lambda (1-z)+
\frac{\langle J\rangle^2}{r_+^{3}}\chi(z) \nonumber\\
&=&\lambda (1-z)+\frac{\langle J\rangle^2}{r_+^{3}}(\chi(0)+z\chi'(0)+\dots)
\label{aw6}
\end{eqnarray}
where in the second line we have expanded $\chi(z)$ about $z=0$. Comparing the coefficient of $z$ on both sides
of this equation, we obtain
\begin{eqnarray}
\frac{\rho}{r_{+}^2}=\lambda-\frac{\langle J\rangle^2}{r_+^{3}}\chi'(0).
\label{aw7}
\end{eqnarray}
Substituting $\chi'(0)$ from eq.(\ref{aw5}) in the above equation, we get
\begin{eqnarray}
\frac{\rho}{r_{+}^2}=\lambda\left\{1+\frac{\langle J\rangle^2}{r_+^{2}}\int_{0}^{1}(1-z)\mathcal{B}(z)dz\right\}.
\label{aw8}
\end{eqnarray}
Finally using eq(s)(\ref{temp}, \ref{lam}) in this equation, we get the following expression for $\langle J\rangle$
\begin{eqnarray}
\langle J\rangle=\gamma T^{2}_{c}\sqrt{1-\frac{T}{T_c}}
\label{aw10}
\end{eqnarray}
where
\begin{eqnarray}
\gamma&=&\frac{2(4\pi/3)^2}{\sqrt{\mathcal{A}}}\\
\mathcal{A}&=&\int_{0}^{1}\frac{z^2 F^{2}(z)}{1+z+z^2}dz~.\nonumber
\label{aw11}
\end{eqnarray}
Computing $\mathcal{A}$ with $\alpha=0.5078$, we get $\gamma\approx123.4$
which is close to the exact numerical result $\gamma\approx104.8$ \cite{hs13a}.
The critical exponent of the expectation value of the condensation operator also comes out to be $1/2$
which is the universal value in the mean field theory.



\section{Critical temperature and condensation values by matching method}
In this section, we shall present a simple analytic treatment of superconductivity and compare our results
with those obtained from the SL approach discussed above. The method involves finding an 
approximate solution around both $z=1$ and $z=0$ using Taylor series expansion 
and then connecting these solutions between $z=1$ and $z=0$ \cite{hs8}.

\subsection{Solution near the horizon: $z=1$}
Near $z=1$, we expand $\phi(z)$ and $\psi(z)$ as 
\begin{eqnarray}
\label{ma1}
\phi(z)&=&\phi(1)-\phi'(1)(1-z)+\frac{1}{2}\phi''(1)(1-z)^2 +...\nonumber\\
&\approx&-\phi'(1)(1-z)+\frac{1}{2}\phi''(1)(1-z)^2\\
\psi(z)&=&\psi(1)-\psi'(1)(1-z)+\frac{1}{2}\psi''(1)(1-z)^2 +...\nonumber\\
&\approx&\psi(1)+\frac{1}{2}\psi''(1)(1-z)^2 
\label{ma8}
\end{eqnarray}
where we have used the boundary condition $\phi(1)=0$ in the second line of eq.(\ref{ma1}) and 
$\psi'(1)=0$ (which readily follows by multiplying 
eq.(\ref{e1a}) by $(1-z^3)^2)$ and then setting $z=1$ and using $f(1)=0$ and $\phi(1)=0$) in the second line of eq.(\ref{ma8}). 
We also set $\phi'(1)<0$ and $\psi(1)>0$ for the positivity of $\phi(z)$ and $\psi(z)$.

\noindent From eq.(\ref{e1aa}), we compute the coefficient of the third term in eq.(\ref{ma1})
\begin{eqnarray}
\phi''(1)=\lim_{z\rightarrow1}\frac{\psi^{2}(z)\phi(z)}{r_{+}^2(1-z^3)}=-\frac{\psi^{2}(1)\phi'(1)}{3r_{+}^2}
\label{ma2}
\end{eqnarray}
Similarly, from eq.(\ref{e1a}), we compute the coefficient of the third term in eq.(\ref{ma8})
\begin{eqnarray}
\psi''(1)&=&\lim_{z\rightarrow1}\left\{\frac{3z^2}{1-z^3}\psi'(z)-\frac{\phi^{2}(z)\psi(z)}{r_{+}^2 (1-z^3)^{2}}\right\}\nonumber\\
&=& -\psi''(1)-\frac{\phi'^{2}(1)\psi(1)}{9r_{+}^2}
\label{ma7}
\end{eqnarray}
which gives
\begin{eqnarray}
\psi''(1)&=&-\frac{\phi'^{2}(1)\psi(1)}{18r_{+}^2}~.
\label{ma7a}
\end{eqnarray}
Setting $\phi'(1)=-\beta$ and $\psi(1)=\alpha$ and substituting eq.(\ref{ma2}) in eq.(\ref{ma1}) 
and eq.(\ref{ma7a}) in eq.(\ref{ma8}), we finally get the solutions for $\phi(z)$ and $\psi(z)$ near $z=1$
\begin{eqnarray}
\label{ma3}
\phi(z)&\approx& \beta(1-z)+\frac{\alpha^2 \beta}{6r_{+}^2}(1-z)^2 \\
\psi(z)&\approx& \alpha-\frac{\alpha\tilde{\beta}^2}{36}(1-z)^2\quad;\quad \tilde\beta=\frac{\beta}{r_+}~.
\label{ma3aa}
\end{eqnarray}

\subsection{Solution near the asymptotic $AdS$ region: $z=0$}
Near $z=0$, we expand $\phi(z)$ and $\psi(z)$ using the asymptotic solutions (\ref{bound1}) and (\ref{bound2}) as
\begin{eqnarray}
\label{bound1a}
\phi(z)&=&\mu-\frac{\rho }{r_+}z+\frac{1}{2}\phi''(0)z^2 +...\\
\psi(z)&=&\frac{\psi^{(1)}}{r_+}z+...
\label{bound2a}
\end{eqnarray}
Now the second derivative of $\phi(z)$ evaluated at $z=0$ is given by
\begin{eqnarray}
\phi''(0)=\lim_{z\rightarrow0}\frac{\psi^{2}(z)\phi(z)}{r_{+}^2(1-z^3)}=0.
\label{ma29}
\end{eqnarray}
Hence we have
\begin{eqnarray}
\label{bound3}
\phi(z)&=&\mu-\frac{\rho }{r_+}z\\
\psi(z)&=&\frac{\psi^{(1)}}{r_+}z = \frac{\langle J\rangle}{\sqrt{2}r_+}z~.
\label{bound4}
\end{eqnarray}


\subsection{Matching and phase transition}
We now connect the solutions (\ref{ma3}), (\ref{ma3aa}) with (\ref{bound3}) and (\ref{bound4}) respectively.
To do this, we take the derivative of the eq.(\ref{ma3}) with respect to $z$ 
and compare with the derivative of eq.(\ref{bound3})
with respect to $z$, which finally yields,
\begin{eqnarray}
\frac{\rho}{r_+}=\beta+\frac{\alpha^2 \beta}{3r_{+}^2}(1-z)~.
\label{ma4}
\end{eqnarray}
Rearranging the above equation and using eq.(\ref{temp}) to write $r_{+}$ in terms of $T$, we get
\begin{eqnarray}
\alpha=\frac{4\pi}{3}\sqrt{\frac{6}{(1-z)}}T_c \left(1-\frac{T}{T_c}\right)^{1/2}
\label{ma5}
\end{eqnarray}
where the critical temperature $T_c$ is given by
\begin{eqnarray}
T_c = \frac{3}{4\pi}\frac{1}{\sqrt{\tilde\beta}}\rho^{1/2}~.
\label{ma6}
\end{eqnarray}
Now comparing eq(s) (\ref{ma3aa}) and (\ref{bound4}), we get
\begin{eqnarray}
\frac{\langle J\rangle}{\sqrt 2 r_+} z=\alpha-\frac{\alpha\tilde{\beta}^2}{36}(1-z)^2.
\label{ma9}
\end{eqnarray}
Differentiating the above equation with respect to $z$, we obtain
\begin{eqnarray}
\frac{\langle J\rangle}{\sqrt 2 r_+} =\frac{\alpha\tilde{\beta}^2}{18}(1-z).
\label{ma10}
\end{eqnarray}
Dividing eq.(\ref{ma9}) by eq.(\ref{ma10}) and rearranging, we get
\begin{eqnarray}
\tilde{{\beta}^2}=\frac{36}{1-z^2}~.
\label{ma11}
\end{eqnarray}
Substituting this expression in eq.(\ref{ma6}) gives
\begin{eqnarray}
T_c = \kappa\rho^{1/2}
\label{ma600}
\end{eqnarray}
where
\begin{eqnarray}
\kappa=\frac{3}{4\pi}\frac{(1-z^2)^{1/4}}{\sqrt{6}}~.
\label{ma601}
\end{eqnarray}
Similarly, substituting the above expression for $\tilde{{\beta}^2}$ in any one of the eq(s)(\ref{ma9})
or (\ref{ma10}) and using eq.(\ref{ma5}), we obtain
\begin{eqnarray}
\langle J\rangle&=&\gamma T T_{c}\left(1-\frac{T}{T_c}\right)^{1/2}\nonumber\\
&\approx&\gamma T_{c}^{2}\left(1-\frac{T}{T_c}\right)^{1/2}
\label{ma12}
\end{eqnarray}
where
\begin{eqnarray}
\gamma=\left(\frac{4\pi}{3}\right)^2 \sqrt{\frac{48}{(1-z)(1+z)^2}}~.
\label{ma12a}
\end{eqnarray}
The above expressions for the critical temperature $T_c$ and the expectation value of the condensation operator 
$\langle J\rangle$ depends on the value of $z$ we choose. Note that $\gamma$ diverges for $z=1$ and therefore it indicates
that the matching of the solutions near the horizon and near the asymptotic $AdS$ region should be done for $z<1$. 

\noindent In the table below (table 1), we list the values of $\kappa$ and $\gamma$ for some values of $z$ in the interval $(0,1)$. 
\begin{table}[htb]
\caption{Values of $\kappa$ and $\gamma$ for various values of $z$ in the interval $(0,1)$}   
\centering                          
\begin{tabular}{c c c c c c c}            
\hline\hline                        
$z$ & $\kappa$  & $\gamma$ & \\ [0.05ex]
\hline
0.1 & 0.0972 & 116.5\\                              
0.2 & 0.0965 & 113.3 \\
0.3 & 0.0952 & 111.8 \\ 
0.4 & 0.0933 & 112.1 \\
0.5 & 0.0907 & 114.6\\
0.6 & 0.0872 & 120.1\\ [0.05ex]  
\hline                              
\end{tabular}\label{E1}  
\end{table}

\noindent We now compare (in the table below (table 2)) our analytical values for $\kappa$ and $\gamma$ obtained by the SL and matching approach with the existing numerical results in the literature \cite{hs13, hs13a}. In order to do that we choose the value of $\kappa_{matching}$ and $\gamma_{matching}$ corresponding to $z=0.3$.
\begin{table}[htb]
\caption{A comparison of the analytical (SL and mathching methods) and numerical results for the critical temperature and the expectation value of the condensation operator}   
\centering                          
\begin{tabular}{c c c c c c c}            
\hline\hline                        
$\kappa_{SL}$ & $\kappa_{matching}$  & $\kappa_{numerical}$ &$\gamma_{SL}$& $\gamma_{matching}$
 & $\gamma_{numerical} $ &  \\ [0.05ex]
\hline
0.1239 & 0.0952 & 0.125 & 123.4 & 111.8 & 104.8   \\ [0.05ex]  
\hline                              
\end{tabular}\label{E1}  
\end{table}

Both from table 1 and table 2, we observe that for $z=0.1$, $\kappa_{matching}$ is closest to both $\kappa_{SL}$ and $\kappa_{numerical}$.
However, on the other hand, $\gamma_{matching}$ is always found to be closer to $\gamma_{numerical}$ compared to $\gamma_{SL}$. For $z=0.3$, we observe that $\gamma_{matching}$ is closest to $\gamma_{numerical}$. Interestingly this does not seems to change the value of  $\kappa_{matching}$ does not change
much from the value at $z=0.1$. As we increase $z$,
both $\kappa_{matching}$ and $\gamma_{matching}$ starts to deviate away from the corresponding numerical values. As $z\rightarrow1$,
the analytical value for $\gamma$ obtained from the matching method diverges which clearly indicates that the matching should be done
closer to $z=0$.

\section{Conclusions}
In this paper, we perform analytic computation of holographic $ p $-wave superconductors in $AdS_{4}$-Schwarzschild background
by two methods, one based on 
the Sturm-Liouville eigenvalue problem and the other based on the matching of the solutions to the field equations near the horizon and near the asymptotic $AdS$ region. The relation between the critical temperature and the charge density has been obtained and the
dependence of the expectation value of the condensation operator on the temperature has been worked out.
Our results are in very good agreement with the existing numerical results \cite{hs13, hs13a}. The critical exponent of the condensation also comes out to be $ 1/2 $  which is the universal value in the mean field theory.

\section*{Acknowledgments} DR would like to thank CSIR for financial support.
SG would like to thank Inter University Centre for Astronomy $\&$ Astrophysics, Pune, India
for providing facilities.


\end{document}